\documentstyle[pra,epsf,aps,twocolumn]{revtex}

\def\cm#1{}

\begin{document}
\title{{
Strong-Coupling
$\phi^4$-Theory
in $4- \epsilon$ Dimensions,
and Critical Exponents
}}
\author{Hagen Kleinert%
 \thanks{Email: kleinert@physik.fu-berlin.de \hfil \newline URL:
http://www.physik.fu-berlin.de/\~{}kleinert \hfil
}}
\address{Institut f\"ur Theoretische Physik,\\
Freie Universit\"at Berlin, Arnimallee 14,
14195 Berlin, Germany}
\maketitle
\begin{abstract}
With the help of variational perturbation theory
we continue the
renormalization constants
$\phi^4$-theories in $4- \epsilon$ dimensions
to strong bare couplings $g_0$ and
find their
power behavior in $g_0$,
thereby
determining
all critical exponents
without renormalization group techniques.
\end{abstract}

%
~\\
\noindent
{\bf 1.}
In a recent paper \cite{sc} we have shown
that there exists a simple way of extracting the
strong-coupling  properties of a $\phi^4$-theory
from perturbation expansions.
In particular, we were able to find power behavior of
the renormalization constants
in the limit of large couplings, and from this all critical exponents.
 of
the system.
By using the known expansion coefficients of the
renormalization constants
in three dimensions up to six loops
we derived extremely accurate
values for the critical exponents.
The method is a systematic
extension
of the Feynman-Kleinert variational approximation to path
 integrals \cite{tk} to arbitrary orders
 \cite{systematic}.
For
an anharmonic oscillator,
the derived  variational perturbation expansions
 converge uniformly
and exponentially fast, like $e^{-{\rm const}\times N^{1/3}}$
in
 the order
$N$
of
   the approximation  \cite{PI,similar}. The same type of
convergence seems to exist also
for the $\phi^4$-theory if the power $1/3$
is replaced by $1- \omega$, where $ \omega$ is the critical exponent governing the approach to
scaling  \cite{sc}.
This exponent plays a crucial role in the development of the theory.

{\bf 2.} Variational perturbation expansions
have the important property of possessing a
good strong-coupling limit as was first shown for the harmonic oscillator
 \cite{JK1,JK2}.
The speed of convergence
turned out to be governed
by the
 convergence
radius
of the strong-coupling expansion \cite{JK3,Guida}.
The good strong-coupling properties
have enabled us
to
set up
a simple
algorithm
 for
deriving
uniformly
convergent
approximations to functions
of which one knows
a few Taylor coefficients
and an important scaling property:
they approach a constant value
with a given inverse power of the variable.
The renormalized coupling constant $g$ of a $\phi^4$-theory has precisely this property
as a function of the bare coupling constant $g_0$.
In $D=4- \epsilon$ dimensions, it approaches a constant value $g^*$
for increasing bare coupling constant $g_0$ like
\begin{equation}
g(g_0)=g^*-\frac{{\rm const}}{g_0^{  \omega/ \epsilon}}+\dots~,
\label{appr}\end{equation}
where $g^*$ is the infrared-stable
fixed point and $ \omega$ is called the critical
exponent of the approach to scaling.

The purpose of this paper is to point out that
the theory developed in \cite{sc}
for a three-dimensional $\phi^4$-theory
can easily be applied in $D=4- \epsilon$ dimensions
with beautiful results at the two-loop level.

{\bf 3.} Let us briefly recall the relevant formulas.
Consider a function $f(g_0)$
for which we know $N$ expansion terms,
$ f_N(g_0)=\sum_{n=0}^N  a_n  g_0  ^n$,
and the fact that
it approaches a constant value $f^*$
in the form of an inverse power series
 $f_M(g_0)=  \sum_{m=0}^M b_m  (g_0 ^{-2/q}) ^m$
with a finite convergence radius $g_s$ (simple examples were treated
in \cite{Interpolation}).
Then the $N$th approximation to the
value $f^*$ is obtained from the formula
\begin{eqnarray}
&&f_N^* =\mathop{\rm opt}_{\hat{g}_0}\left[
\sum_{j=0}^N a_{j}^{\rm } \hat{g}_0^j
 \sum_{k=0}^{N-j}
      \left( \begin{array}{c}
              - q j/2 \\ k
             \end{array}
      \right)
     (-1)^{k}   \right] ,
\label{coeffb}\end{eqnarray}
where the expression in brackets
has to be optimized in the variational parameter
$\hat g_0$.
The optimum is
the smoothest among all real extrema. If there are no real extrema,
the
turning points serve the same purpose.

The derivation  of this expression
is simple:
We replace $g_0$ in $f_N(g_0)$ trivially by
$\bar g_0\equiv g_0/\kappa^q$ with $ \kappa=1$.
Then we rewrite, again trivially, $ \kappa^{-q}$ as $ (K^2+ \kappa^2-K^2)^{-q/2}$
with an arbitrary parameter $K$.
Each term is now expanded in powers of $r=(\kappa^2-K^2)/K^2$
assuming $r$ to be of the order $g_0$.
Taking the limit $ g_0\rightarrow \infty$
at a fixed     ratio
 $\hat g_0\equiv g_0/K^q$,
so that $K\rightarrow \infty$ like $g_0^{1/q}$ and
$r\rightarrow -1$,
we obtain (\ref{coeffb}).
Since the final result to all orders cannot depend on the
arbitrary parameter $K$, we expect the best result to any finite order to
be optimal at an extremal value of $K$, i.e., of $\hat g_0$.

The  strong-coupling approach to the
limiting value
$r= -1+ \kappa^2/K^2
=-1+O(g_0^{-2/q})$
 implies
the leading correction
 to
$f^*_N$ to be
of the order of $g_0^{-2/q} $.
Application of the theory to a function with the
strong-coupling
behavior
(\ref{appr}) requires therefore
a parameter
$q=2 \epsilon/ \omega$
in formula (\ref{coeffb}).

For $N=2$ and $3$ one can give analytic expressions
for the strong-coupling limits
(\ref{coeffb}). Setting $ \rho\equiv 1+q/2=1+ \epsilon/ \omega$, we find for $N=2$
\begin{eqnarray}
f_2^*= \mathop{\rm opt}_{\hat{g}_0}\left[
a_0+
a_1\rho \hat g_0 +
a_2\hat g_0^2
\right] =a_0-\frac{1}{4}\frac{a_1^2}{a_2} \rho^2.
\label{f2@}\end{eqnarray}
For $N=3$, we obtain from the extrema
\begin{eqnarray}
f_3^*&=&  \mathop{\rm opt}_{\hat{g}_0}\left[
a_0+{\scriptstyle\frac{ 1}{ 2}}a_1 \rho( \rho+1)\hat{g}_0+a_2(2 \rho-1)
\hat{g}_0^2
+a_3\hat{g}_0^3 \right]        \nonumber \\
&=&a_0-\frac{1}{3}\frac{\bar a_1\bar a_2}{a_3}\left(1-\frac{2}{3}r
\right)
+\frac{2}{27}\frac{\bar a_2^3}{a_3^2}\left(1-r \right),
\label{f3@}\end{eqnarray}
where
$
r\equiv
 \sqrt{1-3{\bar a_1a_3}/{\bar a_2^2}}$ and
$ \bar a_1 \equiv {\scriptstyle\frac{ 1}{ 2}}a_1 \rho( \rho+1) $
and
$ \bar a_2 \equiv a_2(2 \rho-1) $.
The
positive
square root must be taken
  to connect $g_3^*$ smoothly to
$g_2^*$ in the limit of a vanishing coefficient of $g_0^3$.
If the square root is imaginary, the optimum is given by
the unique turning point, leading once more to (\ref{f3@}) but
with $r=0$.

The parameter $ \rho=1+ \epsilon/ \omega$
can be determined from the expansion coefficients
of a function $F(g_0)$ as follows. Assuming  $F(g_0)$
to be constant $F^*$ in the strong-coupling limit,
the logarithmic derivative $f(g_0)\equiv g_0F'(g_0)/F(g_0)$ must vanish at $g_0=\infty$.
If $F(g_0)$
starts out as $A_0+A_1g_0+\dots$ or $A_1g_0+A_2g_0^2+\dots~$,
the logarithmic derivative is
\begin{eqnarray}
f(g_0)&=& A'_1 g_0+(2 A'_2- A'_1{}^2)g_0^2
\nonumber \\&&+
(A'_1{}^3-3 A'_1 A'_2 +3 A'_3)g_0^3
+\dots~,
\label{omscal}\end{eqnarray}
where $ A'_i=A_i/A_0$, or
\begin{eqnarray}
f(g_0)&=&1+\hat A_2 g_0+(2\hat A_3-\hat A_2^2)g_0^2
\nonumber \\&&+
(\hat A_2^3-3\hat A_2\hat A_3+3\hat A_4)g_0^3
+\dots~,
\label{omscal2}\end{eqnarray}
where $\hat A_i=A_i/A_1$.
The expansion coefficients on the right-hand sides
are then inserted into
(\ref{f2@}) or (\ref{f3@}), and the left-hand sides have to vanish
to ensure that $F(g_0)\rightarrow F^*$.

If the approach $F(g_0)\rightarrow F^*$
is of the type
(\ref{appr}),
the function
\begin{eqnarray}
h(g_0)&\equiv& g_0\frac{F''(g_0)}{F'(g_0)} =2\hat A_2 g_0
+(-4\hat A^2_2+6\hat A_3)g_0^2
\nonumber \\&&
+(8\hat A_2^3-18\hat A_2 \hat A_3 +12\hat A_4)g_0^3
+\dots~
\label{extraeq}\end{eqnarray}
must have the strong-coupling limit
\begin{equation}
h(g_0)\rightarrow h^*=-\frac{ \omega} \epsilon-1.
\label{fome@}\end{equation}

{\bf 4.}
These formulas
are now applied
to the
renormalization constants of the
$\phi^4$-theory in $D=4- \epsilon$ dimensions
with the bare euclidean action
\begin{equation}
{\cal A}\!=\!
\int\! d^Dx\left\{ \frac{1}{2}\left[\partial \phi_0(x)\right]^2\!
\!+\!   \frac{1}{2} m_0^2\phi_0^2(x)
\!+\!(4\pi)^2\frac{  \lambda_0}{4!}\left[  \phi_0^2(x)\right] ^2\!\right\}\!.
\label{@}\end{equation}
The field $\phi_0(x)$ is an $n$-dimensional vector, and the action is
O($n$)-symmetric in this vector space.
The Ising model corresponds to $n=1$,
the critical behavior of percolation is described by $n=0$,
superfluid phase transitions by $n=2$, and classical Heisenberg
magnetic
systems by $n=3$.

By calculating the Feynman integrals
regularized via an expansion in $ \epsilon=4-D$
with the help of an
arbitrary mass scale $\mu$,
one obtains renormalized values
of
mass, coupling constant, and field related to the bare input
quantities by
renormalization constants
$Z_{\phi},Z_{m},Z_{g}$:
\begin{equation}
m_0^2
=m^2\ Z_{m}Z^{-1}_{\phi},~~
   \lambda_0=  \lambda\,{Z_g}Z^{-2}_{\phi}, ~~
\phi_0=\phi\,\,Z_{\phi}^{1/2}.
\label{gefung}
\label{ren@}\end{equation}
Up to two loops,
perturbation theory yields
the following
expansions in powers of the dimensionless reduced
coupling constant
$g\equiv  \lambda/\mu^ \epsilon$:
\begin{eqnarray}
{ g}&=&{ g}_0 -
\frac{ n+8}{3 \epsilon }  g_0^2 +
  \,\left[ \frac{(n+8)^2}{9 \epsilon^2}+\frac{3n+14}{6 \epsilon}\right]
{ g_0^{3}}
, \label{gequ@}\\
\frac{m^2}{m_0^2}&=&1-
\frac{n+2}{3}\frac{g_0}{ \epsilon}
+\frac{n+2}{9}\left[\frac{n+5}{ \epsilon^2}+\frac{5}{4 \epsilon}
\right]
{g_0^2},\label{gfg-0}\\
\frac{\phi^2}{\phi_0^2}&=&1+\frac{n+2}{36}\frac{g_0^2}{ \epsilon}
.
\label{gfg-1}\end{eqnarray}
We now
set the scale parameter $\mu$ equal to $m$
and consider all quantities as functions
of
 $g_0= \lambda/m^ \epsilon$.
In order to describe second-order phase transitions,
we let $m_0^2$ go to zero like $\tau =$const$\times( T-T_c)$
as the temperature
$T$ approaches the critical temperature $T_c$.
Then also $m^2$ will go to zero,
and thus
$g_0$
to infinity.
Assuming the theory to scale as suggested by experiments,
we now determine the value
of the renormalized coupling constant $g$
in the strong-coupling limit $g_0\rightarrow \infty$,
and the approach to it, assuming the behavior
(\ref{appr}). For this we apply formula (\ref{coeffb}) to
$g(g_0)$ to find $g^*$,
and use the vanishing of (\ref{omscal2}),
or (\ref{fome@}) with (\ref{extraeq}) at strong couplings to determine $ \omega$.
Under the scaling assumption,
the ratios
$m^2/m_0^2$ and $\phi^2/\phi_0^2$ have the limiting power behavior
for small $m$:
\begin{equation}
\frac{m^2}{m_0^2}\propto  g_0^{- \eta_m/ \epsilon}\propto m^{ \eta_m/ \epsilon},~~~~~~
\frac{\phi^2}{\phi_0^2}\propto  g_0^{ \eta/ \epsilon}\propto m^{ -  \eta/ \epsilon}.
\label{phimass}\end{equation}
The powers can therefore be calculated
from the strong-coupling limits of the
logarithmic derivatives
\begin{eqnarray}
  \eta_{m}({ g}_0)\!=\!
- \epsilon\frac{d}{d\log  g_0}\log\frac{m^2}{m^2_0},~
  \eta({ g_0})
\!=\!
 \epsilon\frac{d}{d\log  g_0}\log\frac{\phi^2}{\phi^2_0}
 .\!\!\label{etame2}
\label{etaetam@}\end{eqnarray}
From (\ref{gfg-0}) and (\ref{gfg-1})
find the expansions
\begin{eqnarray}
 \eta _m(g_0)&=&\frac{n+2}{3 }g_0-\frac{n+2}{18}\left(5+2\frac{n+8} \epsilon\right)g_0^2,\label{@etam}\\
 \eta(g_0)&=&\frac{n+2}{18}g_0^2.
\label{@eta}\end{eqnarray}

When approaching the
second-order phase transitions,
where $m_0^2$
vanishes like $\tau \equiv (T-T_c)$,  $m^2$ vanishes
 with a different power of $\tau $.
This power is obtained from the first equation
in (\ref{phimass}) which shows that
$m\propto \tau ^{1/(2- \eta_m)}$. Experiments observe
that the coherence length of fluctuations $\xi=1/m$ increases near $T_c$
like $\tau ^{- \nu}$, so that
we derive for the critical exponent
$ \nu$ a value $1/(2- \eta_m)$.
Similarly we see
from the first equation in
(\ref{phimass}) that
 the scaling dimension $D/2-1$ of the free field  $\phi_0$
for $T\rightarrow T_c$
is changed, in the strong-coupling limit $g_0\rightarrow \infty $,
to $D/2-1+ \eta/2$, the number $ \eta$ being the so-called anomalous dimension of the field.
This implies a change in the large-distance behavior
of the correlation functions  $\langle \phi(x)\phi(0)\rangle$
at $T_c$ from the free-field behavior
 $r^{-D+2}$ to
 $r^{-D+2- \eta}$.
The magnetic susceptibility is determined by
the integrated correlation
function $\langle \phi_0(x)\phi_0(0)\rangle$.
At zero coupling constant $g_0$, this is proportional to $1/m_0^2\propto \tau^{-1} $,
which is changed by
 fluctuations to $m^{-2}\phi_0^2/\phi^2$. This has a temperature
behavior $m^{-(2- \eta)}=\tau ^{ -\nu(2- \eta)}\equiv \tau ^{- \gamma}$, which
 defines the critical exponent $ \gamma= \nu(2- \eta)$ observable
in magnetic experiments.
Using
$ \nu=1/(2- \eta_m)$ and the expansions
(\ref{@etam}), (\ref{@eta}),
we obtain for $ \gamma(g_0)$
the perturbation expansion up to second order in  $g_0$:
\begin{eqnarray}
  \gamma(g_0)&=&1+\frac{n+2}{6}{ g}_0+ \frac{ n+2 }{36}\left(n-4-2\frac{n+8}{ \epsilon}\right)
 g_0^2
.
\label{gamma}\end{eqnarray}

All calculations in this note will be restricted to
the two loop level, which will
be sufficient to demonstrate the power
and beauty of
the new strong-coupling theory
with analytical
results.

{\bf 5.} We begin by calculating
the critical exponent $ \omega$
from the
requirement that $g(g_0)$ has a constant
strong-coupling limit,
implying the vanishing of (\ref{omscal2}) for $g_0\rightarrow \infty$.
From the expansion
(\ref{gequ@}) we obtain a logarithmic derivative
(\ref{omscal2}) up to the term $g_0^2$, so that
Eq.~(\ref{f2@}) can be used to
find the scaling condition
\begin{equation}
0=1-\frac{1}{4}\frac{\hat A_2^2}{2\hat A_3-\hat A_2^2} \rho^2
\label{ex@}\end{equation}
This gives
\begin{equation}
 \rho= \sqrt{8\hat A_3/\hat A_2^2-4}.
\label{rho@}\end{equation}
Since $ \omega$ must be greater than zero, only
the positive square root is physical.
With the explicit coefficients
$A_1,A_2,A_3$ of expansion (\ref{gequ@}), this becomes
\begin{equation}
 \rho=
2\sqrt{1+{3}\frac{3n+14}{(n+8)^2} \epsilon}  .
\label{x22@}\end{equation}
The associated critical exponent $ \omega= \epsilon/( \rho-1)$
is plotted in
Fig.~1.
It has the $ \epsilon$-expansion
\begin{equation}
  \omega= \epsilon  -3\frac{3n+14}{(n+8)^2} \epsilon^2+\dots~,
\label{omep@}\end{equation}
which is also shown in Fig. 1, and
agrees with the first two terms obtained from
renormalization group calculations \cite{10}.

From Eqs.~(\ref{fome@}), (\ref{extraeq}), and (\ref{f2@}) we obtain
 for the critical exponent $  \omega$ a further equation
\begin{eqnarray}
-\frac{ \omega}{ \epsilon}-1= -\frac{ \rho}{ \rho-1}&=&
-\frac{1}{2}\frac{\hat A_2^2 \, \rho^2  }{3\hat A_3-2\hat A_2^2}
.
\label{@}\end{eqnarray}
which is solved by
\begin{equation}
 \rho=\frac{1}{2}+ \sqrt{\frac{6\hat A_3}{\hat A_2^2}-\frac{15}{4}},
\label{@}\end{equation}
with the positive sign of the square root
ensuring
a positive
$ \omega$.
Inserting the coefficients of
(\ref{gequ@}), this becomes
\begin{equation}
 \rho=\frac{1}{2}+ \frac{3}{2} \sqrt{1+4\frac{3n+14}{(n+8)^2} \epsilon}.
\label{sign}\end{equation}
The associated critical exponent $ \omega= \epsilon/( \rho-1)$
has
the same $ \epsilon$-expansion (\ref{omep@})
as the previous approximation (\ref{x22@}).
The full approximations based on (\ref{sign}) is indistinguishable from
the earlier one in the plot
of Fig. 1.

Having determined $ \omega$, we can now calculate $g^*$.
Inserting the first two coefficients of the expansion
(\ref{gequ@}) into (\ref{f2@}) we obtain
\begin{equation}
g^*_2= a_0
-
\frac{1}{4}\frac{a_1^2}{a_2} \rho^2.
\label{@}\end{equation}
Inserting (\ref{x22@}), this yields
\begin{equation}
g^*_2=\frac{3}{n+8} \epsilon+
9   \frac{3n+14}{(n+8)^3} \epsilon^2,
\label{epsex}\end{equation}
which is precisely the well-known $ \epsilon$-expansion
of $g^*$ in renormalization group calculations up to the second order.
Including the next coefficient,
we can use formula (\ref{f3@})
to calculate
the next approximation $g_3^*$. At $ \epsilon=1$, the square root turns out to be
imaginary, so that it has to be omitted (corresponding
to the
turning point as optimum).
The resulting curve lies slightly ($\approx 8\%$) above
the curve (\ref{epsex}), i.e., represents a worse approximation
than
(\ref{epsex}).  Indeed, the $ \epsilon^3$-term in $g_3^*$ is
$81(3n+14)^2/8(n+8)^5$ and
disagrees in sign with the exact term
 $
{ \epsilon^3}[{3}\left(
 -33n^3  +110n^2+1760n +  4544\right)/8
$ $-36\zeta(3)(n+8)\left(5n+22\right)]/{(n+8)^5},$
%
which we would find by calculating $ \rho$
from an expansion (\ref{gequ@}) with one more power in $g_0$.

We now turn to the critical exponent
$ \nu$.
Taking the expansion (\ref{@etam}) to $g_0\rightarrow \infty$,
we obtain from
formula (\ref{f2@}) the limiting value
\begin{equation}
 \eta_m= \frac{\epsilon}4 \frac{n+2}{n+8+5 \epsilon/2} \rho^2.
\label{@}\end{equation}
%
%
The corresponding
$ \nu=1/(2- \eta_m)$
is plotted in Fig. 1.
With the approximation (\ref{x22@}) for $ \rho$
we find for $ \nu$ the $ \epsilon$-expansion
\begin{equation}
 \nu=\frac{1}{2}+\frac{1}{4}\frac{n+2}{n+8} \epsilon+\frac{(n+2)(n+3)(n+20)}{8(n+8)^3} \epsilon^2+\dots~,
\label{@}\end{equation}
which is also shown in Fir.~1, and agrees with
renormalization group results to this order.

As a third independent
critical exponent we calculate
$ \gamma=(2- \eta)/(2- \eta_m)$ by inserting
the coefficients of the expansion (\ref{gamma})
into formula~(\ref{f2@}), which yields
\begin{equation}
  \gamma={1}+\frac{ \epsilon}{8}\frac{n+2}{n+8-(n-4)\epsilon/{2}} \rho^2 ,
\label{@}\end{equation}
plotted in Fig.~1. This has an $ \epsilon$-expansion
\begin{equation}
 \gamma={1}+\frac{1}{2}\frac{n+2}{n+8} \epsilon
+\frac{1}{4}\frac{(n+2)(n^2+22n+52)}{(n+8)^3} \epsilon^2+\dots~,
\label{@}\end{equation}
shown again in Fig.~1, and
agreeing with renormalization group results to this order.
The full approximation
is plotted in Fig. 1.
The critical exponent
 $ \eta=
2- \gamma/ \nu$ has the $ \epsilon$-expansion
%
$ \eta={(n+2)} \epsilon^2/2{(n+8)^2}+\dots~.  $
%

{\bf 6.}
We conclude that variational
strong-coupling theory
can easily
be applied to
 $\phi^4$-theories
in $D=4- \epsilon$ dimensions
and yields resummed expressions
for the $ \epsilon$-dependence
of all  critical exponents.
Their $ \epsilon$-expansions agree with
those obtained form renormalization group
calculations.
The calculations here are supposed to illustrate the new calculational
procedure and do not yet give good results for the critical exponent
$ \omega$.
In order to achieve high accuracy, we shall have to go
to five loops and incorporate knowledge of
the large-order behavior of the expansion coefficients.

Recent results by A. Pelissetto and E. Vicari (University of Pisa preprint IFUP-TH 52/97)
on the renormalization in O($n$)-models via $1/n$ expansions
will lead to further improvements of the theory.

~~\\~~\\
{\bf Acknowledgment}~\\~\\
The author thanks Dr. V. Schulte-Frohlinde for useful discussions.

%
%

%
%
%
 \pagebreak~\\
\begin{figure}[tbhp]
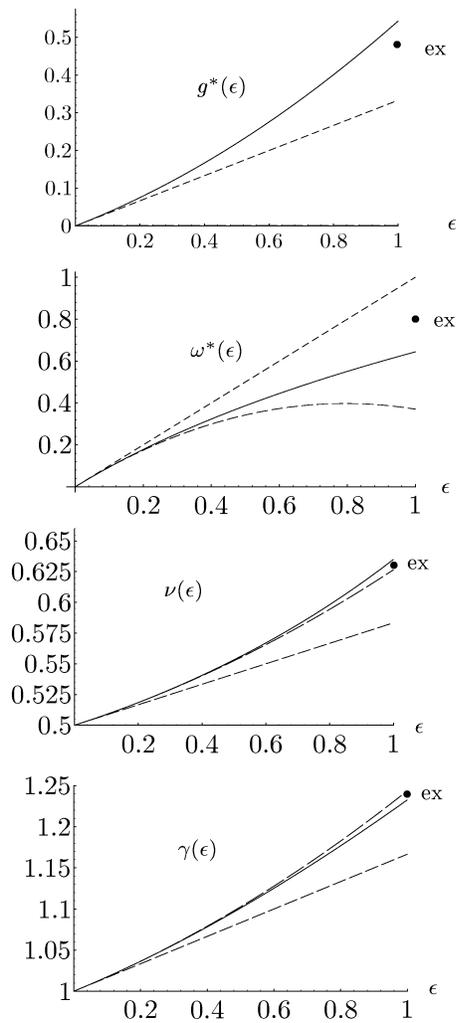

\input all.tps   ~\\
\caption[]{
For the Ising universality class ($n=1$),
the first figure shows the renormalized coupling
at infinite bare coupling
as a function of $ \epsilon=4-D$ calculated via variational perturbation theory
from the first two perturbative
expansion terms. The curve coincides with the $ \epsilon$-expansion
up to order $ \epsilon^2$.
The dashed curve
indicates
the linear term.
The other figures show the critical exponents $ \omega$, $ \nu$, and $ \gamma$.
Dashed curves indicate linear and quadratic $ \epsilon$-expansions.
The dots mark presently accepted values
of $g^*\approx0.48\pm 0.003$, $ \omega\approx0.802\pm0.003$,
$ \nu=0.630\pm0.002$, and $ \gamma=1.241\pm0.004$
obtained from six-loop calculations \cite{sc}.
}
\label{omf}\end{figure}
~\\[-1.8cm]


\begin{thebibliography}{11}
\bibitem{sc}
   H. Kleinert, APS E-Print aps1997jun25\_001 .

\bibitem{tk}
     R.P Feynman and H. Kleinert,
     Phys.\ Rev.\  {\bf  A34}, 5080 (1986).
A similar approach has been
pursued independently by \\
R.~Giachetti and
V.~Tognetti,
Phys.\ Rev.\ Lett.~{\bf 55}, 912 (1985);
Int.\ J.~Magn.\ Mater.~{\bf 54-57}, 861 (1986);
{R.~Giachetti},
{V.~Tognetti}, and
{R.~Vaia}, Phys.
Rev.~{\bf B33}, 7647 (1986).

\bibitem{systematic}
   H. Kleinert,
     Phys.\ Lett. {\bf A173}, 332 (1993).
 \bibitem{PI}
  H. Kleinert,
     {\em Path Integrals in Quantum Mechanics,
     Statistics and Polymer Physics,\/}
     World Scientific, Singapore 1995.
\bibitem{similar}
For earlier similar
expansions
see
by\\
R. Seznec and J. Zinn-Justin, J. Math. Phys. {\bf 20}, 1398 (1979);
{T.~Barnes and G.I.~Ghandour}, Phys.\ Rev.\ {\bf D22},
  924 (1980);
{B.S.~Shaverdyan} and
{A.G.~Usherveridze},
 Phys.\ Lett.\ {\bf B123}, 316 (1983);
{P.M.~Stevenson}, Phys.\ Rev.\ {\bf D30}, 1712 (1985);
{\bf D32}, 1389 (1985);
{P.M.~Stevenson} and
{ R.~Tarrach}, Phys.\ Lett.\ {\bf B176}, 436 (1986);
{A.~Okopinska}, Phys.\ Rev. {\bf D35}, 1835 (1987);
{\bf D36}, 2415 (1987);
{W.~Namgung},
{P.M.~Stevenson}, and
{J.F.~Reed},
 Z.~Phys.\ {\bf C45}, 47 (1989);
{U.~Ritschel}, Phys.\ Lett.\ {\bf B227}, 44 (1989);
    Z.~Phys.\ {\bf C51}, 469 (1991);
{M.H.~Thoma}, Z.~Phys.\ {\bf C44}, 343 (1991);
{I.~Stancu} and
{P.M.~Stevenson},
  Phys.\ Rev.\ {\bf D42}, 2710 (1991);
{R.~Tarrach}, Phys.\ Lett.\ {\bf B262}, 294 (1991);
{H.~Haugerud} and
{F.~Raunda}, Phys.\ Rev.\ {\bf D43},
  2736 (1991);
{A.N.~Sissakian},
{I.L.~Solivtosv}, and
  {O.Y.~Sheychenko},
  Phys.\ Lett.\ {\bf B313}, 367 (1993);
{A.~Duncan} and
{H.F.~Jones}, Phys.~Rev.~ {\bf D47}, 2560
    (1993);
\bibitem{JK1}
  W.~Janke and H.~Kleinert,
  Phys.\ Lett.~{\bf A199}, 287 (1995).
 \bibitem{JK2}
  W.~Janke and H.~Kleinert,
   Phys.\ Rev.\ Lett.\ {\bf 75}, 2787 (1995).
   (quant-ph/9502019). \\
  That paper contains references to earlier, less
accurate
calculations of strong-coupling expansion coefficients
from weak-coupling perturbation theory, in particular\\
F.M. Fern\'{a}ndez and R. Guardiola, J. Phys. {\bf A26}, 7169 (1993);
F.M. Fern\'{a}ndez, Phys. Lett. {\bf A166}, 173 (1992);
R. Guardiola, M.A. Sol\'{\i}s, and J. Ros, Nuovo Cimento {\bf B107},
713 (1992).
A.V. Turbiner and A.G. Ushveridze, J. Math. Phys. {\bf 29}, 2053 (1988);
B. Bonnier, M. Hontebeyrie, and E.H. Ticembal, J. Math. Phys. {\bf 26}, 3048
(1985); Those works did not extract the exponential law of convergence
 from their data.



\bibitem{JK3}
 H.~Kleinert and W.~Janke,
Phys. Lett. {\bf A206}, 283 (1995) (quant-ph/9509005).



\bibitem{Guida}
A convergence proof which is completely
equivalent our results in \cite{JK3} was given by \\
R. Guida, K. Konishi, and H. Suzuki,
 Annals Phys. {\bf 249}, 109 (1996)
(hep-th/9505084).\\
Predecessors of these works
which
did not explain
the
exponentially fast convergence in the strong-couplings
limit observed
in Ref. \cite{JK2}
are\\
I.R.C. Buckley, A. Duncan, and H.F. Jones, Phys. Rev. {\bf D47}, 2554 (1993);
C.M. Bender, A. Duncan, and H.F. Jones, Phys. Rev. {\bf D49}, 4219 (1994);
A. Duncan and H.F. Jones, Phys. Rev. {\bf D47}, 2560 (1993);
C. Arvanitis, H.F. Jones, and C.S. Parker,
Phys. Rev. {\bf D52}, 3704 (1995) (hep-th/9502386);
R. Guida, K. Konishi, and H. Suzuki,
Annals Phys. {\bf 241}, 152 (1995) (hep-th/9407027).




\bibitem{Interpolation}
 H.~Kleinert, Phys.\ Lett.\ {\bf A207}, 133 (1995)  (quant-ph/9507005).
\bibitem{10}
H. Kleinert, J. Neu, V. Schulte-Frohlinde, K.G. Chetyrkin, and S.A. Larin,
Phys. Lett. {\bf B272}, 39 (1991) (hep-th/9503230);
H. Kleinert and V. Schulte-Frohlinde,
Phys. Lett. {\bf B342}, 284 (1995) (cond-mat/9503038).





\end{thebibliography}
\end{document}